\documentclass[preprint,aps,preprintnumbers,nofootinbib,superscriptaddress,showpacs]{revtex4}
\usepackage{amsmath,amssymb,amsbsy}
\usepackage{graphicx}
\usepackage{epsfig}

\newcommand*{\ie}{\textit{i.e.},\ }

\newcommand{\Delstar}{\ensuremath{\Delta^{\raise0.18ex\hbox{${\scriptstyle *}$}}}}
\def\gtwid{{\,\raise.35ex\hbox{$>$\kern-.75em\lower1ex\hbox{$\sim$}}\,}}
\def\ltwid{{\,\raise.35ex\hbox{$<$\kern-.75em\lower1ex\hbox{$\sim$}}\,}}
\def\leftvec{{\raise1.5ex\hbox{$\leftarrow$}\kern-1.00em}}
\def\rightvec{{\raise1.5ex\hbox{$\rightarrow$}\kern-1.00em}}
\def\half{{\scriptstyle \raise.2ex\hbox{${1\over2}$}}}
\def\threehalves{{\scriptstyle \raise.15ex\hbox{${3\over2}$}}}
\def\third{{\scriptstyle \raise.15ex\hbox{${1\over3}$}}}
\def\third{{\scriptstyle \raise.15ex\hbox{${1\over3}$}}}
\def\twothirds{{\scriptstyle \raise.15ex\hbox{${2\over3}$}}}
\def\fourth{{\scriptstyle \raise.15ex\hbox{${1\over4}$}}}

\newcommand{\dbar}{\ensuremath{\overline{d}}}

\newcommand*{\bea}{\begin{eqnarray}}
\newcommand*{\eea}{\end{eqnarray}}
\newcommand*{\be}{\begin{equation}}
\newcommand*{\ee}{\end{equation}}

\newcommand*{\CPT}{\raise0.4ex\hbox{$\chi$}PT}
\newcommand*{\chpt}{\raise0.4ex\hbox{$\chi$}PT}
\newcommand*{\schpt}{S\raise0.4ex\hbox{$\chi$}PT}

\def\eqref#1{{(\ref{#1})}}

\def\CO{{\cal O}}

\def\bar{\overline}

\def\bk{$B_K\;$}

\def\bea{\begin{eqnarray}}
\def\eea{\end{eqnarray}}

\def\spose#1{\hbox to 0pt{#1\hss}}
\def\ltapprox{\mathrel{\spose{\lower 3pt\hbox{$\mathchar"218$}}
 \raise 2.0pt\hbox{$\mathchar"13C$}}}
\def\gtapprox{\mathrel{\spose{\lower 3pt\hbox{$\mathchar"218$}}
 \raise 2.0pt\hbox{$\mathchar"13E$}}}
\def\inapprox{\mathrel{\spose{\lower 3pt\hbox{$\mathchar"218$}}
 \raise 2.0pt\hbox{$\mathchar"232$}}}

\begin{document}

\preprint{FERMILAB-PUB-08-053-T}

\vphantom{}

\title{Discretization effects and the scalar meson correlator\\ in mixed-action lattice simulations}

\author{C.\ Aubin}
\email[]{caaubin@wm.edu}
\affiliation{Department of Physics, College of William and Mary, Williamsburg, VA 23187}

\author{Jack Laiho}
\email[]{jlaiho@fnal.gov}
\affiliation{Physics Department, Washington University, St. Louis, MO 63130}

\author{Ruth S. Van de Water}
\email[]{ruthv@fnal.gov}
\affiliation{Theoretical Physics Department, Fermilab, Batavia, IL 60510}

\date{\today}

%=================================================
%The abstract
%=================================================
\begin{abstract}
We study discretization effects in a mixed-action lattice theory with domain-wall valence quarks and Asqtad-improved staggered sea quarks.  At the level of the chiral effective Lagrangian, discretization effects in the mixed-action theory give rise to two new parameters as compared to the lowest order Lagrangian for rooted staggered fermions -- the residual quark mass, $m_\textrm{res}$, and the mixed valence-sea meson mass-splitting, $\Delta_\textrm{mix}$.  We find that $m_\textrm{res}$, which parameterizes explicit chiral symmetry breaking in the mixed-action theory, is approximately a quarter the size of our lightest valence quark mass on our coarser lattice spacing, and of comparable size to that of simulations by the RBC and UKQCD Collaborations.  We also find that the size of $\Delta_\textrm{mix}$ is comparable to the size of the smallest of the staggered meson taste-splittings measured by the MILC Collaboration.  Because lattice artifacts are different in the valence and sea sectors of the mixed-action theory, they give rise to unitarity-violating effects that disappear in the continuum limit, some of which should be described by mixed-action chiral perturbation theory (MA\chpt).  Such effects are expected to be mild for many quantities of interest, but are expected to be significant in the case of the isovector scalar ($a_0$) correlator.  Specifically, once the parameters  $m_\textrm{res}$,  $\Delta_\textrm{mix}$, and two others that can be determined from the light pseudoscalar meson spectrum are known, the two-particle intermediate state ``bubble" contribution to the scalar correlator is completely predicted within MA\chpt.  We find that the behavior of the scalar meson correlator is quantitatively consistent with the MA\chpt\ prediction;  this supports the claim that MA\chpt\ describes the dominant unitarity-violating effects in the mixed-action theory and can therefore be used to remove lattice artifacts and recover physical quantities.
\end{abstract}

\maketitle

%=================================================
\section{Introduction}
\label{sec:Intro}
%=================================================

Advances in both computers and algorithms now allow realistic nonperturbative calculations of hadron masses and matrix elements from QCD first principles.  In particular, the results of lattice simulations with three flavors of improved staggered dynamical quarks are in good agreement with experimental measurements for a wide range of light meson, heavy-light meson, and heavy-heavy meson quantities~\cite{Davies:2003ik}.  These successes, which include both ``postdictions" such as the pion and kaon decay constants~\cite{Aubin:2004fs} and predictions such as the shape of the $D\to K\ell \nu$ form factor~\cite{Aubin:2004ej}, indicate that many of the systematic uncertainties associated with lattice QCD calculations are now under control.  Therefore lattice QCD can now be used to calculate reliably many weak matrix elements that cannot be determined experimentally but are important for flavor physics phenomenology.   These matrix elements are needed to interpret the results of 
experiments such as those at the Tevatron, $B$-factories, and, in the near future, the LHC,
and for precise tests of the Standard Model in the quark flavor sector.  

This paper is the first in a series leading to the calculation of the kaon bag parameter~\cite{Aubin:2007pt}, $B_K$, with all sources of systematic error under control using the mixed-action lattice QCD simulation scheme pioneered by the LHP Collaboration~\cite{Renner:2004ck}.  $B_K$ parameterizes the hadronic part of neutral kaon mixing, and, when combined with an experimental measurement of $\epsilon_K$, constrains the apex of the  Cabibbo-Kobayashi-Maskawa unitarity triangle.  Because $\epsilon_K$ has been well-measured experimentally~\cite{Yao:2006px}, the dominant uncertainty in this constraint is that of $B_K$.  It is likely that whatever new physics exists has additional $CP$-violating phases which will manifest themselves as inconsistencies between measurements that are predicted to be identical within the Standard Model.  Thus a precise lattice determination of $B_K$ helps constrain physics beyond the Standard Model and is an important goal of the phenomenology community.  The mixed-action method, which we describe in detail in the following section, employs domain-wall valence quarks and improved staggered sea quarks, and is particularly well-suited for the numerical calculation of weak matrix elements such as $B_K$. 

In order to understand and demonstrate control over the various sources of systematic error that enter the numerical calculation of $B_K$, we have divided this project into smaller pieces that are both interesting by themselves, and, when viewed as a whole, will lend credibility to our $B_K$ calculation.  Because discretization errors will be one of the primary sources of uncertainty in $B_K$, in this first paper we study their effects in the mixed-action theory.  In particular, we quantify the size of the explicit chiral symmetry breaking parameter, $m_\textrm{res}$, and the mixed-action parameter, $\Delta_\textrm{mix}$.  We also test the range of applicability of mixed-action Chiral Perturbation Theory (MA\chpt) by analyzing numerical mixed-action data for the isovector scalar correlator, in which unitarity-violating discretization effects are expected to be substantial for our choice of simulation parameters.  In the second paper we will present results for the pseudoscalar decay constants, $f_\pi$ and $f_K$.  Because $f_\pi$ is well-known experimentally, it provides a test of the mixed-action lattice methodology, especially the combined chiral-continuum extrapolation using MA\chpt.  The ratio $f_K/f_\pi$ then allows a model-independent determination of the CKM matrix element $V_{us}$~\cite{Marciano:2004uf}.  Only after we have tested the mixed-action method using the known quantity $f_\pi$ will we present results for $B_K$.  These results will appear in a subsequent publication.

\bigskip

In this work we study discretization effects in the mixed-action theory.  We calculate the residual quark mass, $m_\textrm{res}$.  We obtain a value of $m_\textrm{res} \approx 2.7$ MeV in the chiral limit on the $a \approx 0.12$ fm (``coarse") MILC lattices; $m_\textrm{res}$ is even smaller, approximately 1 MeV, on the $a \approx 0.09$ fm (``fine") MILC lattices. Given these values, we find that $m_\textrm{res}$ on the coarse lattices is a quarter the size of our smallest valence quark mass, and comparable to that of the RBC and UKQCD Collaborations on a similar lattice spacing~\cite{Antonio:2007tr}, indicating that the amount of chiral symmetry breaking is acceptably small in our mixed-action simulations.  We also calculate the mixed valence-sea meson splitting, $\Delta_\textrm{mix}$, on both the coarse and fine lattices.  We find that $a^2 \Delta_\textrm{mix} \approx (280 \textrm{MeV})^2$ on the coarse lattices and $a^2 \Delta_\textrm{mix} \approx (190 \textrm{ MeV})^2$ on the fine lattices, which is about a factor of two smaller.  Thus these results are consistent with the expected $\CO(a^2)$ scaling behavior of  $a^2 \Delta_\textrm{mix}$, as well as the independent determination on the coarse lattices by Orginos and Walker-Loud \cite{Orginos:2007tw}.  We note that the numerical values of both $m_\textrm{res}$ and $\Delta_\textrm{mix}$ are unique to this particular choice of mixed-action scheme, and would be different if, for example, one were to use highly-improved staggered quark (HISQ) gauge configurations for the sea sector~\cite{Follana:2007uv}.

One might be concerned that the use of a mixed action could introduce new theoretical complications, and, in particular, sizeable unitarity-violating effects that are not described by mixed-action Chiral Perturbation Theory (MA\chpt).  Because the valence and sea quarks in mixed-action simulations generically have different discretization effects, one cannot tune the valence-valence and sea-sea pseudoscalar mesons to remove all unitarity violations at nonzero lattice spacing.  In other words, the mixed-action lattice theory is always partially quenched, and one cannot recover full QCD until the continuum limit.  Thus, in order to extract physical quantities from mixed-action simulations, one must be able to account for all sources of lattice discretization effects (at a given order in MA\chpt) and remove them.  In quantities such as $f_K$ and $B_K$, unitarity-violating discretization effects are relatively mild, and only appear as meson mass-shifts inside chiral logarithms.  One would like to demonstrate, however, that even when the unitarity-violation is more pronounced, it can still be described by MA\chpt.  

In this work we show that  MA\chpt\ accurately describes the behavior of the isovector scalar 2-point correlation function.  The $a_0$ correlator is particularly sensitive to unitarity-violating discretization effects in the mixed-action theory because it receives contributions from flavor-neutral two-meson intermediate states.  The ``bubble" contribution to the scalar correlator was calculated in MA\chpt\ by Prelovsek~\cite{Prelovsek:2005rf}.  The expression contains only a few low-energy constants, all of which can be determined in fits to pseudoscalar meson mass data.  Thus the size and shape of the bubble contribution to the $a_0$ correlator is completely predicted within MA\chpt.  We find that, for all valence-sea mass combinations on both the coarse and fine lattices, the predicted bubble contribution is quantitatively consistent with our mixed-action 2-point data.  Thus we conclude that MA\chpt\ describes the dominant unitarity-violating effects in the numerical simulations.  In the case of the scalar, we cannot actually determine the $a_0$ mass with our existing data because, for most of our coarse lattice data, the size of the bubble contribution swamps the leading exponential contribution from the $a_0$ meson over almost the entire time range.  In the case of $B_K$, however, unitarity-violating non-analytic discretization effects are predicted to be less than a percent of the continuum value of $B_K$ over the relevant extrapolation range on the coarse MILC lattices~\cite{Aubin:2006hg}.  Thus we can use MA\chpt\ to remove these effects and recover precisely the continuum value of $B_K$.

\bigskip
   
This paper is organized as follows.  In Section~\ref{sec:MA_Sims}, we describe the actions and parameters used in our lattice calculations.  We calculate the residual quark mass, $m_\textrm{res}$, in Section~\ref{sec:mres} and the mixed-meson splitting, $\Delta_\textrm{mix}$, in Section~\ref{sec:DelMix}.  Both of these quantities appear in MA\chpt\  expressions, and we use our value of $\Delta_\textrm{mix}$ to analyze the behavior of the $a_0$ correlator within the framework of MA\chpt\ in Section~\ref{sec:Scalar}.  In Section~\ref{sec:Conc} we summarize our results and conclude.

%=================================================
\section{Mixed-action lattice simulations}
\label{sec:MA_Sims}
%=================================================

In this section we motivate the use of mixed-action lattice simulations for the determination of hadronic observables, and, in particular, weak matrix elements.  We then describe the details of the lattice actions and parameters used in our numerical mixed-action calculations.

\subsection{Theoretical advantages}

Any numerical lattice simulation in which the actions for the valence and sea quarks are different is a mixed-action calculation.  In this work, however, we use the terminology ``mixed-action" specifically to refer to the choice of actions used by the LHP Collaboration, \ie a domain-wall valence quark action on top of a staggered sea.  

Staggered fermions are computationally cheap as compared to other standard discretizations.  Thus they allow the lightest dynamical quark masses and finest lattice spacings currently available in lattice simulations~\cite{Bernard:2007ps}.  This numerical affordability comes at the cost, however, of additional theoretical complications.  Staggered quarks come in four degenerate species called tastes~\cite{Susskind:1976jm}; consequently each flavor of staggered meson exists in sixteen tastes.  Although the sixteen light pseudoscalar meson tastes are degenerate in the continuum, the symmetry that relates them is broken at nonzero lattice spacing, leading to discretization errors that are of $\CO(a^2)$~\cite{Lee:1999zx}.  Because these errors are numerically significant at current lattice spacings, staggered lattice data must be fit to functional forms that include taste-breaking discretization effects calculated in staggered Chiral Perturbation Theory (S\chpt)~\cite{Lee:1999zx,Aubin:2003mg, Aubin:2003uc, Sharpe:2004is}.  The unphysical effects can then be subtracted to recover the desired physical quantity.\footnote{Although there is no rigorous proof, there is considerable supporting evidence that the rooting procedure used in numerical simulations with staggered quarks is correct.  We therefore work under the plausible assumption that the continuum limit of the rooted staggered theory is QCD.  Both numerical and theoretical evidence supporting the rooting procedure are reviewed by Sharpe in Ref.~\cite{SharpePlenary} and by Kronfeld in Ref.~\cite{Kronfeld:2007ek}.}  
The staggered chiral effective Lagrangian contains a number of low-energy constants in addition to those of continuum \chpt.  In many quantities of interest such as masses and decay constants, these new parameters primarily enter next-to-leading order (NLO) chiral expressions in a straightforward way by producing an additive shift to the meson masses that appear inside chiral logarithms~\cite{Aubin:2003mg, Aubin:2003uc}.  In other quantities (such as $B_K$) that are not protected by a partially-conserved current, however, taste-symmetry breaking leads to a more complicated operator mixing pattern that must either be accounted for with fully nonperturbative operator renormalization, which is difficult, or with the inclusion of many new operators in the S\chpt\ calculation, which introduces many new undetermined coefficients in the joint chiral-continuum extrapolation~\cite{VandeWater:2005uq}.

The calculation of weak matrix elements such as \bk with
domain-wall quarks~\cite{Kaplan:1992bt, Shamir:1993zy}, on the other hand,
is theoretically simpler than that with staggered quarks because domain-wall quarks do not occur in multiple species.  Furthermore, they retain exact chiral symmetry at nonzero lattice spacing up to exponentially small corrections, the size of which can be controlled by the length of the 5th dimension~\cite{Antonio:2007tr}.  Consequently, while the $\Delta S=2$ lattice operator still mixes with wrong-chirality operators,  there are significantly fewer operators than in the staggered case, and nonperturbative renormalization can be applied more easily~\cite{Aoki:2007xm}.  Lattice simulations with domain-wall fermions, however, are computationally more expensive than those with staggered fermions with comparable masses and lattice spacings~\cite{Clark:2006wq}.

The LHP mixed-action scheme retains the primary advantage of staggered lattice simulations while reducing the complications associated with taste-symmetry breaking.  Thus it is particularly well-suited for the numerical calculation of $B_K$.  The mixed-action method allows for light dynamical quark masses and fine lattice spacings, but at the computational cost of a quenched domain-wall lattice simulation.  The chiral symmetry of the domain-wall valence quarks eliminates mixing with operators of other tastes and minimizes mixing with those of other chiralities, thereby making the nonperturbative renormalization more tractable.  Furthermore, it makes NLO mixed-action \chpt\ expressions, including that for $B_K$, largely continuum-like~\cite{Bar:2005tu,Tiburzi:2005is,Chen:2005ab,Chen:2007ug,Aubin:2006hg}.  

\subsection{Numerical details}

We now describe the details of our numerical mixed-action simulations.  A summary of the ensembles and valence quark masses used in this work can be found in Table~\ref{tab:MILC_ens}.

\begin{table}
\caption{Parameters of the MILC improved staggered gauge configurations and domain-wall valence quark propagators used in this work.  Columns three and four show the nominal up/down ($m_l$) and strange quark ($m_s$) masses in the sea, along with the corresponding pseudoscalar taste pion mass.  Columns five and six list our partially-quenched valence quark masses, along with our lightest available domain-wall pion mass.}
\label{tab:MILC_ens}
\begin{tabular}{lllcccccc} \\ \hline

\multicolumn{4}{c}{\qquad} & \multicolumn{3}{c}{sea sector} & \multicolumn{2}{c}{valence sector} \\[-0.5mm]
$a$(fm) && L  && $a m_l / am_s$ & $a m_\pi$ & \qquad &  $am_{v}$ & $a m_\pi$   \\[0.5mm] \hline\hline

0.09 && 28 && 0.0062/0.031 & 0.14789(18) && 0.0062, 0.0124, 0.0186, 0.046 & 0.1201(14) \\
0.09 && 28 && 0.0124/0.031 & 0.20635(18) && 0.0062, 0.0124, 0.0186, 0.046 & 0.1213(15) \\
\hline

0.12 && 24 && 0.005/0.05 & 0.15971(20) & \qquad\qquad & 0.007, 0.02, 0.03, 0.05, 0.065 & 0.1733(12) \\
0.12 && 20 & \qquad\qquad & 0.007/0.05 & 0.18891(20) && 0.01, 0.02, 0.03, 0.04, 0.05, 0.065 & 0.1971(11) \\
0.12 && 20 && 0.01/0.05 & 0.22447(17) && 0.01, 0.02, 0.03, 0.05, 0.065 & 0.19931(94) \\
0.12 && 20 && 0.02/0.05 & 0.31125(16) && 0.01, 0.03, 0.05, 0.065 & 0.1938(21) \\
\hline

\end{tabular}\end{table}

\bigskip

We use the MILC lattices with 2+1 dynamical flavors of Asqtad-improved staggered fermions~\cite{Susskind:1976jm}; a detailed description of the simulation parameters can be found in Refs.~\cite{Bernard:2001av,Aubin:2004wf}.    We have data on both the ``coarse" ($a\approx 0.12$ fm) and ``fine" ($a\approx 0.09$ fm) MILC ensembles, which have physical volumes ranging from approximately (2.5 -- 3 fm)${}^3$.    For each ensemble, the mass of the dynamical strange quark is close to its physical value, and the masses of the up and down quarks are degenerate.  The ratios of the nominal up/down quark mass to the nominal strange quark mass in the sea sector are given in the the third column of Table~\ref{tab:MILC_ens}.

We generate domain-wall valence quark propagators~\cite{Kaplan:1992bt,Shamir:1993zy} using the Chroma software system for lattice QCD~\cite{Edwards:2004sx}.  Like the LHP Collaboration~\cite{Renner:2004ck}, we HYP-smear the lattices using the standard parameters given in Ref.~\cite{Hasenfratz:2001hp} in order to reduce the size of explicit chiral symmetry breaking and proximity to the Aoki phase~\cite{Aoki:1983qi}.  We also use the same values for the domain-wall height, $M_5$=1.7, and extent of the fifth dimension, $L_S$=16.  Using these parameters, we find that the residual quark mass is acceptably small; details of our $m_\textrm{res}$ calculation are given in Section~\ref{sec:mres}. 

We choose bare domain-wall quark masses in order to best suit the numerical calculation of $B_K$, although we are also using the resulting propagators for quantities such as the pseudoscalar decay constants.  We therefore need a wide range of valence quark masses that allows us both to extrapolate our lattice data to the physical up/down quark mass and to interpolate to the physical strange quark mass.  In order to keep finite-volume effects under control, however, we restrict the quantity $m_\pi L \gtapprox 4$.  Based upon these constraints, we have chosen the domain-wall valence quark masses listed in the fifth column of Table~\ref{tab:MILC_ens}.  Note that, unlike in other mixed staggered sea, domain-wall valence lattice simulations we have not made any attempt to tune the domain-wall pion mass in the valence sector to match the lightest staggered pion mass in the sea sector.  This is because, although many possible tunings exist, none completely removes unitarity-violating effects from the mixed-action theory at nonzero lattice spacing; full QCD is only recovered in the continuum limit.  We have, instead, generated many partially-quenched points, and will use the appropriate expressions in mixed-action Chiral Perturbation Theory~\cite{Bar:2005tu,Aubin:2006hg} to extrapolate our pseudoscalar decay constant and $B_K$ data in future work.  

We have generated Coulomb gauge-fixed wall-source propagators with periodic and antiperiodic boundary conditions starting at timeslice zero for use in the calculation of the $K^0-\bar{K^0}$ matrix element.  In order to make best use of the available computing time, we have only generated quark propagators on every fourth MILC gauge configuration;  thus all of our data points are relatively uncorrelated.  These wall-source propagators are used for our determination of the residual quark mass, $m_\textrm{res}$, in Section~\ref{sec:mres}.  In order to compute the parameter $\Delta_\textrm{mix}$, we formed mixed valence-sea mesons by tying the wall-source propagators together with Coulomb gauge-fixed (but not HYP-smeared) point source Asqtad valence quark propagators made with the MILC code~\cite{MILCCode}.  The results for $\Delta_\textrm{mix}$ are given in Section~\ref{sec:DelMix}.  We have also generated periodic-boundary-condition Landau gauge-fixed point-source propagators for nonperturbative operator renormalization using the method of Rome-Southampton~\cite{Martinelli:1994ty}.  These propagators were used for our preliminary analysis of the isovector scalar 2-point correlation function in Ref.~\cite{Aubin:2007wr} because a point source is required for comparison to the MA\chpt\ expression in Ref.~\cite{Prelovsek:2005rf}.  It turned out, however, that relatively large statistical errors in the point-point scalar data made it difficult to resolve the small bubble contribution on the fine lattices.   We therefore generated additional Coulomb gauge-fixed $Z_2$ random-wall source propagators on a subset of our ensembles in order to reduce the size of the statistical errors in the scalar correlator data.  The random-wall source gives the same normalization as a point source (up to a factor of the spatial volume), but averages over all of the spatial points on the source timeslice.  The analysis of the $a_0$ correlator with the improved data is presented in Section~\ref{sec:Scalar}.    

In order to convert dimensionful quantities determined in our mixed-action lattice simulations into physical units, we use the MILC Collaboration's recent determination of the scale $r_1 = 0.318(7)$ fm~\cite{Bernard:2007ps}.  The Sommer scale-setting method~\cite{Sommer:1993ce} has the advantage that the ratio $r_1/a$ can be determined precisely from a fit to the static quark potential~\cite{Bernard:2000gd,Aubin:2004wf}.  By converting all of our data from lattice spacing units into $r_1$ units before performing any chiral fits, we account for slight differences in the value of the lattice spacing between ensembles.  MILC obtains the absolute scale $r_1$ by first using the HPQCD Collaboration's calculation of $a^{-1}$ on the coarse and fine lattices from Upsilon spectroscopy~\cite{Gray:2005ur} to get $r_1$ on the coarse and fine lattices, and then extrapolating these values to the continuum.

%=================================================
%\section{Chiral symmetry breaking and $m_\textrm{res}$}
\section{The residual quark mass}
\label{sec:mres}
%=================================================

In this section we determine the value of the chiral symmetry breaking parameter, $m_{res}$, in our mixed-action simulations.  

\bigskip

Domain-wall fermions respect flavor symmetry and have an approximate chiral symmetry \cite{Kaplan:1992bt, Shamir:1993zy, Furman:1994ky}.  This is possible because of the introduction of an extra dimension;  the Dirac operator for domain-wall fermions is a five-dimensional operator, with free boundary conditions in the fifth dimension.  Light chiral fermion modes are exponentially bound to the four-dimensional surfaces at the ends of the fifth dimension.  In the limit that the separation between the two domain walls, commonly denoted $L_s$, is taken to infinity, the chiral symmetry becomes exact.  In practice the value of $L_s$ must be finite in numerical simulations, typically between 10-20 lattice spacing units, and chiral symmetry is not exactly maintained \cite{Antonio:2007tr}.  

As in the case of Wilson fermions, the leading chiral symmetry breaking terms are proportional to the lattice spacing $a$.  However, for Wilson fermions this term is ${\cal O}(1)$, whereas for domain-wall fermions this term is ${\cal O}(10^{-3})$ for typical choices of $L_s$.  The only effect of this leading-order term is a shift of the effective bare quark mass by a small residual mass, $m_{\textrm{res}}$.  The generic, weak-coupling behavior of chiral symmetry breaking leads to the expectation that $m_{\textrm{res}}$ depends on $L_s$ as $a \exp(-\alpha L_s)$, for some constant $\alpha$ \cite{Golterman:2004mf}.\footnote{An analysis of the domain-wall transfer matrix reveals that there is also a contribution to the residual quark mass that depends upon $L_s$ as $\rho(0)/L_s$, where $\rho(0)$ is the density of near-zero eigenmodes~\cite{Golterman:2003qe}.  The size of both the exponential and power-law contributions to $m_{\textrm{res}}$ have been measured by the RBC and UKQCD Collaborations for various domain-wall sea quark actions at multiple lattice spacings~\cite{Antonio:2007tr}.  The relative importance of the power-law contribution increases with $L_s$ and decreases rapidly with the lattice spacing. In particular, RBC/UKQCD found this term to be comparable to the exponential contribution for $L_s=16$ at $a^{-1} \sim 1.6$ GeV, and 6 times smaller for $L_s=16$ and $a^{-1} \sim 2.3$ GeV.  These lattice spacings are similar to those used in this work.}  Additional chiral symmetry breaking effects are suppressed by higher powers of the lattice spacing, on top of the exponential in $L_s$ suppression factor.  Therefore the leading corrections to domain-wall fermions are ${\cal O}(a \exp(-\alpha L_s))$ and ${\cal O}(a^2)$ \cite{Allton:2007hx}.  

In practice, the residual mass is measured in lattice simulations using the ratio of the midpoint current to the axial current \cite{Blum:2000kn, Aoki:2002vt}: 
\bea
 \label{eq:mres}   R(t)=\frac{\langle \sum_{\vec{x}} J^{a}_{5q}(\vec{x},t)\pi^a(0) \rangle}{\langle \sum_{\vec{x}}J^a_5(\vec{x},t) \pi^a(0) \rangle}.
\eea
As $t\to\infty$, this quantity should approach a constant which is then defined as the residual mass, that is, $R(t)\to m_{\rm res}$.
The axial current in Eq.~(\ref{eq:mres}) is
\bea 
	J^a_5(x)= -\bar{\Psi}(x,L_s-1)P_L t^a \Psi(x,0) + \bar{\Psi}(x,0)P_R t^a \Psi(x,L_s-1),
\eea
and the midpoint current is
\bea 
J^a_{5q}(x)= -\bar{\Psi}(x,L_s/2-1)P_L t^a \Psi(x,L_s/2) + \bar{\Psi}(x,L_s/2)P_R t^a \Psi(x,L_s/2-1),
\eea
where $\Psi$ is the domain-wall field, the flavor matrices $t^a$ are normalized by $\textrm{Tr}(t^a t^b)=\delta^{ab}$, and $P_{R,L}=(1\pm \gamma_5)/2$.

\begin{figure}
\begin{center}
\includegraphics[scale=.55]{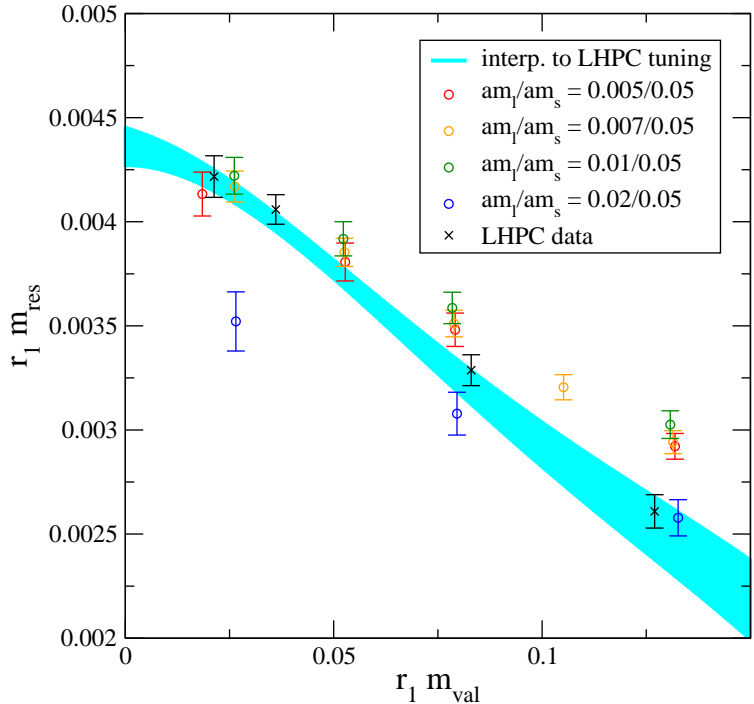}
\caption{Chiral extrapolation of $m_{\textrm{res}}$ on the coarse lattices.
 The curve shows the extrapolation/interpolation
 for points where the domain-wall pion mass is tuned to equal the lightest
 (taste pseudoscalar) staggered pion mass. For comparison, we show the
 determination of $m_{\textrm{res}}$ by the LHP Collaboration, which uses this tuning
 \cite{Renner:2004ck}. \label{fig:mres}}
\end{center}
\end{figure}

When defined as in Eq.~(\ref{eq:mres}), $m_{\textrm{res}}$ is constant up to discretization errors which give $m_{\textrm{res}}$ a dependence on the light quark masses (both valence and sea) and the lattice spacing.  This light quark mass dependence has been observed in a number of earlier simulations.  In our simulations, we follow the method of the LHP Collaboration by applying HYP smearing to the gauge fields before inverting the domain-wall propagators  \cite{Renner:2004ck}.  This smearing reduces the coupling of the domain-wall quarks to high momentum gluons which exchange chirality, and thereby reduces the size of $m_{\textrm{res}}$ on the coarse MILC lattices to an acceptable level.  

In Fig.~\ref{fig:mres} we show our calculated results for $m_{\textrm{res}}$ (as determined from Eq.~(\ref{eq:mres})) for a number of different valence and sea quark values on the coarse MILC lattices. Because the mixed-action theory does not have a full QCD (unitary) point, the curve shows an extrapolation/interpolation to points where the domain-wall pion mass is tuned to equal the lightest (taste pseudoscalar) staggered pion mass.  For comparison, we show the determination of $m_{\textrm{res}}$ by the LHP Collaboration, which uses this tuning~\cite{Renner:2004ck}; 
as one can see, the agreement is good.  The fit function used was a simple polynomial in the valence and sea quark masses, with terms up to fourth order in the valence mass, and linear in the sea mass.  The correlated fit has a C.L. of around $10\%$, which is not unreasonable, given that chiral logarithms are expected to enter at the same order as the linear terms.  We quote a value for $r_1 m_{\textrm{res}}$ in the chiral limit on the coarse ensembles of 0.0044(1)(4), where the first error is statistical, and the second is a systematic error that comes from varying the functional form and mass range used in the fit.  In physical units, $m_{\textrm{res}}$ is around $2.7$ MeV, close to the size of the physical light quark mass.  

\begin{figure}
\begin{center}
\includegraphics[scale=.55]{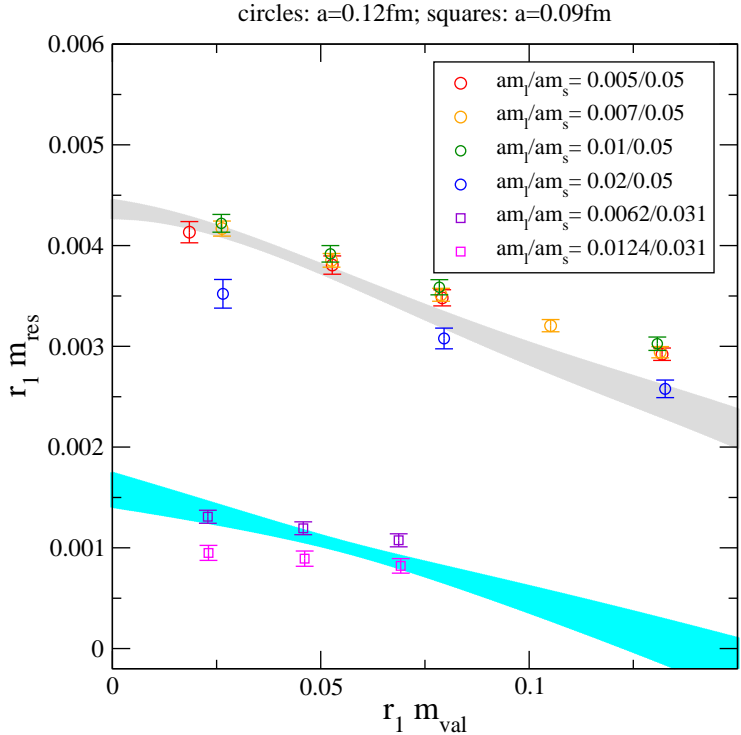}
\caption{Comparison of $r_1 m_{\textrm{res}}$ on the coarse ($a\approx 0.12$ fm) and fine ($a\approx 0.09$ fm) MILC lattices.  The curves are the interpolation/extrapolation to the LHPC tuning.  The value of $r_1 m_{\textrm{res}}$ on the fine lattices is approximately three times smaller than it is on the coarse lattices. \label{fig:mresboth}}
\end{center}
\end{figure}

In Fig.~\ref{fig:mresboth} we compare the values of $m_{\textrm{res}}$ computed on the fine MILC lattices with the results on the coarse lattices.  We observe a significant dependence on the lattice spacing, and $m_{\textrm{res}}$ decreases by a factor of three from the coarse to fine lattices.  The RBC and UKQCD Collaborations see a similar reduction in $m_{\textrm{res}}$ when comparing its value on lattices with nearly the same scales as the coarse and fine MILC lattices, but using three flavors of domain-wall sea quarks instead of improved staggered sea quarks.  We quote a value for $r_1 m_{\textrm{res}}$ in the chiral limit on the fine MILC lattices of 0.0016(2), or approximately $1$ MeV in physical units, where the error is statistical only.

%=================================================
%\section{Taste symmetry breaking and $\Delta_\textrm{mix}$}
\section{The mixed meson mass-splitting}
\label{sec:DelMix}
%=================================================

We have shown that the dominant effect of the residual chiral symmetry breaking is small and under control given the size of the residual quark mass, especially on the finer lattices.  In this section and the following section we demonstrate that discretization effects arising due to the interaction of the valence and sea sectors can be understood and controlled using mixed-action chiral perturbation theory (MA\chpt).  

\bigskip

MA\chpt\ for Ginsparg-Wilson type quarks on a staggered sea was first considered in Ref.~\cite{Bar:2005tu}, where the one-loop formulas for $f_\pi$ and $m_\pi^2$ were derived.  Many additional quantities have since been calculated to one loop in MA\chpt\ \cite{Tiburzi:2005is,Chen:2005ab,Chen:2007ug,Aubin:2006hg}.  It has been demonstrated that there is only one low-energy constant particular to the mixed-action chiral effective theory that appears at one loop.  Additionally, only one of the many parameters coming from the purely staggered sector enters mixed-action expressions at one loop.  Thus only two additional parameters enter the mixed-action chiral formulas as compared to those for purely domain-wall on domain-wall simulations.   These two new parameters are both meson mass-splittings, and one has already been determined by the MILC Collaboration from spectrum calculations so that it could be used as an input to their chiral fits to more complicated quantities \cite{Aubin:2004fs}.  We have calculated the parameter unique to the mixed-action case so that we can use this parameter as an input to our own chiral formulas in the same manner as the MILC Collaboration.

In a mixed-action theory one can have mixed mesons made up of one valence and one sea quark, in addition to mesons made up of two valence or two sea quarks.  We review the tree-level mass relations for these mesons as given in Ref.~\cite{Bar:2005tu}, since they are useful in understanding the leading-order lattice-spacing contributions to mixed-action numerical simulations.  The valence-valence meson mass is given by
\bea 
	m_{vv'}^2 & = & \mu_{\rm dw}(m_v + m_{v'} + 2m_{\rm res})\ ,
\eea 
where we have included the contribution from the residual quark mass.  In the limit that $L_s$ becomes infinite, $m_\textrm{res} \to 0$ and we recover the continuum relation.  Because of taste-symmetry breaking, the staggered theory has 16 pions instead of one.  The tree-level relation for staggered mesons is \cite{Aubin:2003mg}
\bea  
	m_{ss',t}^2 & = & \mu_{\rm stag} (m_s + m_{s'}) + a^2 \Delta_t\ ,
\eea
where the $a^2 \Delta_t$ are the taste-splittings of the 16 taste pions.    For staggered quarks there exists a residual SO(4) taste symmetry at ${\cal O}(a^2)$, such that there is some degeneracy between the 16 pions, and we treat the taste index $t$ as running over the multiplets $P, A, T, V, I$ with degeneracies 1, 4, 6, 4, 1.  The splitting $a^2 \Delta_P$ vanishes because there is an exact (non-singlet) lattice axial symmetry.  It is convenient to express our formulas in terms of the bare lattice masses which enter the lattice calculations, not the renormalized masses.  Since the quark masses using different actions are 
renormalized according to different schemes, we absorb this scheme dependence into separate coefficients, $\mu_{\rm dw}$ and $\mu_{\rm stag}$.  Finally, the mass of a mixed valence-sea meson is 

\bea  
	m_{vs}^2 &=& \mu_{\rm dw}(m_v+m_{\rm res}) + \mu_{\rm stag} m_s + 
	a^2\Delta_{\rm mix} ,   
\eea
where $\Delta_{\rm mix}$ is the only new constant that appears in the mixed-action theory to next-to-leading order.

%The Asqtad improvement of the staggered action reduces the taste splittings in the sea sector to ${\cal O}(\alpha_s^2a^2)$, though these splittings are still quite large on the coarse MILC lattices \cite{Aubin:2004fs}.  This scaling has been confirmed by MILC in their numerical simulations over four different lattice spacings \cite{Bernard:2006wx, Bernard:2007ps}.  

%In order to understand the expected scaling of the $\Delta_{\rm mix}$ term, it is useful to consider a mixed-action theory with two different actions, both of which satisfy the Ginsparg-Wilson relation.  In this case one would still have a $\Delta_{\rm mix}$ term in the tree-level relation for a mixed meson, although the valence-valence and sea-sea relations would be the same as in the continuum.  This splitting in the mixed meson would not necessarily be a short-distance effect (such as staggered taste-symmetry breaking), and thus one would not expect $\Delta_{\rm mix}$ to be suppressed by additional powers of $\alpha_s$.  The same is true in our case, though we will now show that, because $a^2\Delta_{\rm mix}$ is smaller than most of the MILC taste-splittings on the coarse lattices, this relatively weaker scaling behavior is not a concern.

\begin{figure}
\begin{center}
\includegraphics[scale=.55]{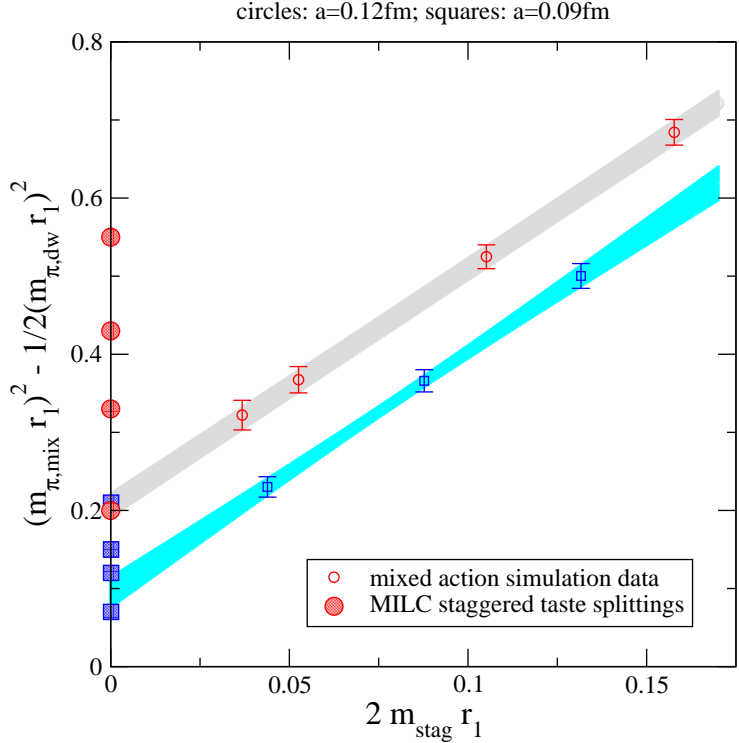}
\caption{Determination of the mixed-action parameter $\Delta_{\textrm{mix}}$ on the coarse and fine MILC lattices.  The vertical axis is a carefully chosen linear combination of squared meson masses, the left side of Eq.~(\ref{eq:DelMix}), such that a linear extrapolation of this quantity in the staggered quark mass gives the parameter $\Delta_{\textrm{mix}}$ as the y-intercept.  The small diamonds (squares) are the coarse (fine) data.  For comparison, the large diamonds (squares), show the values of the staggered taste splittings measured by MILC on the coarse (fine) lattices.   \label{fig:DelMix}}
\end{center}
\end{figure}

We have calculated $\Delta_{\rm mix}$ by computing the mass of a pion made of one domain-wall and one staggered quark.  In order to extract $\Delta_{\rm mix}$ from our spectrum data, it is convenient to construct the following linear combination of squared meson masses:
\bea 
\label{eq:DelMix}
 m^2_{vs} - \frac12 m^2_{vv} = \mu_{\textrm{stag}}m_{s} + a^2 \Delta_{\textrm{mix}}.
	\eea
\noindent We perform a linear fit to the left-hand side of this equation as a function of the staggered mass.  The result is shown in Figure~\ref{fig:DelMix}; in the chiral limit this fit gives $a^2 \Delta_{\rm mix}$.  Figure~\ref{fig:DelMix} also compares our results for $a^2 \Delta_{\rm mix}$ with the four taste-breaking staggered meson splittings as determined by MILC on both the coarse and fine lattices; we find that $\Delta_{\rm mix}$ is around the same size as the smallest of the MILC splittings.  The value of $a^2 \Delta_{\rm mix}$ decreases by roughly a factor of two from coarse to fine lattice spacings, which is consistent with the expected $\CO(a^2)$ scaling behavior.  We quote a value for $r_1^2 a^2 \Delta_{\rm mix}$ on the coarse MILC lattices of 0.207(16), which is $\approx (280 \  \textrm{MeV})^2$ in physical units, and is consistent with the value found in Ref.~\cite{Orginos:2007tw}.  On the fine lattices, we find $r_1^2 a^2 \Delta_{\rm mix}=0.095(20)$, which corresponds to $\approx (190 \  \textrm{MeV})^2$ in physical units.
	
%=================================================
\section{Unitarity violation and the scalar correlator}
\label{sec:Scalar}
%=================================================

Because mixed-action lattice simulations have different discretization errors in the valence and sea sectors, they are not unitary at non-zero lattice spacing.  Such effects are expected to vanish in the continuum limit.  One must take them into account, however, when extracting physical quantities from numerical mixed-action lattice data calculated at fixed lattice spacings.  This can be done using \chpt\ that is formulated for the specific mixed action being used.   In the case of domain-wall valence on staggered sea lattice simulations, for many quantities of interest such as pseudoscalar meson masses, decay constants, and $B_K$, the violation of unitarity appears mildly in chiral expressions as chiral logarithms of the form $m_{\pi,\rm sea}^2 \ln (m_{\pi,\rm val}^2)$~\cite{Bar:2005tu,Aubin:2006hg}.  One would like a strong check, however, that MA\chpt\ truly describes all unitarity-violating effects that are observable in the lattice data, even when the effects are more pronounced.  The behavior of the isovector scalar ($a_0$) correlator provides such a check.  

Quantities such as the $a_0$ correlator and $\pi-\pi$ scattering in the $I=0$ channel are particularly sensitive to unitarity-violating effects due to the presence of flavor-neutral intermediate states~\cite{Bardeen:2001jm,Prelovsek:2005rf}. These give rise to diagrams that are disconnected at the quark level, \ie hairpin diagrams, in which partial quenching effects are more pronounced.  In particular, for the $a_0$ meson, such quark-disconnected diagrams affect the lattice correlator itself.  While our results in the next subsection are specific to this particular choice of mixed action for valence and sea quarks, the issues that we discuss are generic to all mixed-action simulations.  Similar analyses of the $a_0$ correlator have been done for the two-flavor domain-wall case in Ref.~\cite{Prelovsek:2004jp} and for the staggered valence and sea case in Ref.~\cite{Bernard:2007qf}.

\subsection{The $a_0$ in mixed-action \chpt}\label{sec:scalarMA}

\begin{figure}[t]
\begin{center}
\includegraphics[width=6in]{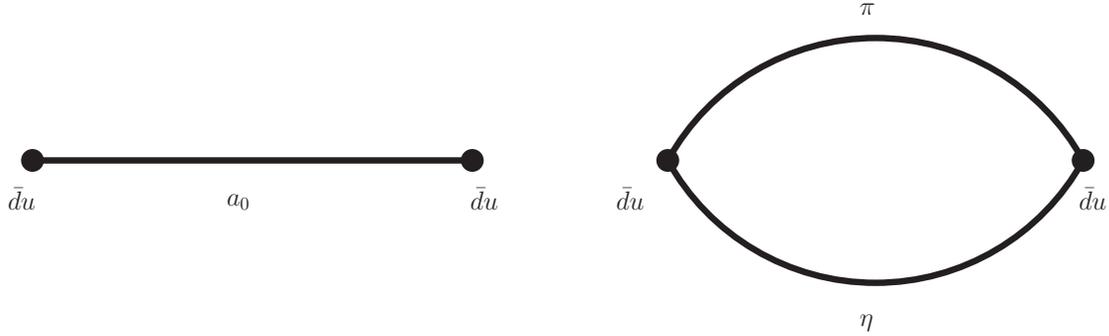}
\caption{Leading contributions to the scalar current.  The first diagram corresponds to the propagation of an $a_0$ meson and the second is one of three ``bubble'' diagrams.  While this figure shows a $\pi$ and an $\eta$ propagating, there are contributions from $\pi-\pi$ and $K-\overline{K}$ intermediate states as well.\label{fig:bubble}}
\end{center}
\end{figure}

The isovector scalar is created with the following local operator at the quark level:
\begin{equation}
	S(\mathbf{x},t) = \dbar(\mathbf{x},t) u(\mathbf{x},t)\ .
\end{equation}
In the chiral effective theory that describes the pseudo-Goldstone boson sector, this is represented by the point current
\begin{equation}
	S_\chi(x) = \mu \left[\Phi^2(x)\right]_{ud}\ .
\label{eq:ChPT_cur}
\end{equation}
In numerical lattice simulations, the $a_0$ correlator is given by $C(t) = \sum_{\mathbf{x}}\left\langle 0|S(\mathbf{x},t)S^\dag(\mathbf{0},0)|0\right\rangle$.
The leading contribution to $C(t)$ comes from the propagation of an $a_0$ meson from time 0 to time $t$.   The propagation of the $a_0$  cannot be handled within \chpt, since in that framework only the light pseudoscalar bosons are dynamical degrees of freedom.  In \chpt\ the scalar propagator leads to a contact term, but for the purposes of our fits, it can be parameterized by an exponential of the form $A\ \textrm{exp}[-m_{a_0}t]$.  It was first noticed in the quenched case, however, that the isovector correlator also receives sizeable contributions from two-particle intermediate states~\cite{Bardeen:2001jm}.  These two-particle ``bubble" contributions, unlike the direct exponential term, can be calculated in \chpt\ using the expression for the scalar current, Eq.~(\ref{eq:ChPT_cur});  an example diagram is shown in Figure~\ref{fig:bubble}.

Thus the leading expression for the lattice $a_0$ correlator has two terms:
\begin{equation}\label{eq:corr}
	C(t) = A e^{-m_{a_0} t } + B(t) + \cdots
\end{equation} 
where the $\cdots$ represent excited state contributions. The bubble term $B(t)$ has been calculated by Prelovsek using one-loop mixed-action \chpt\ in Ref.~\cite{Prelovsek:2005rf}.   The result for 2+1 flavors of sea quarks, taking the time direction to be infinite in length but accounting for the finite spatial extent, is\footnote{The form of $B(t)$ shown here is found by integrating the expression in Ref.~\cite{Prelovsek:2005rf} over the time direction as well as performing the necessary Fourier transform.}
\begin{eqnarray}\label{eq:bubble}
	B(t)
%%%%
	&=& \frac{\mu^2}{3L^3}\sum_{\mathbf{k}}
	\Biggl[
	\frac{2}{9}\frac{e^{-(\omega_{vv} + \omega_{\eta_I}) t}}
	{\omega_{vv}\omega_{\eta_I}}
	\frac{(m_{S_I}^2 - m_{U_I}^2)^2}
	{(m^2_{vv} - m^2_{\eta_I})^2}
	-\frac{e^{-2\omega_{vv} t}}{\omega_{vv}^2}
	\left[
	\frac{3m^2_{vv}(m^2_{vv} - 2m^2_{\eta_I})
	+ 2m^4_{S_I} + m^4_{U_I}}
	{3(m^2_{\eta_I} - m^2_{vv})^2 }\right]
	\nonumber\\&&{}
	- \frac{e^{-2\omega_{vv} t}}{2\omega_{vv}^4}
	\left(\omega_{vv} t + 1\right)
	\frac{(m^2_{U_I} - m^2_{vv})
	(m^2_{S_I} - m^2_{vv} )}{m^2_{\eta_I} - m^2_{vv}}
	+ \frac{3}{2}\frac{e^{-2\omega_{vu} t}}{\omega_{vu}^2}
	+ \frac{3}{4}\frac{e^{-2\omega_{vs} t}}{\omega_{vs}^2}
	\Biggr]\ ,
\end{eqnarray}
where $\omega_i^2 \equiv \sqrt{\mathbf{k}^2 + m_i^2}$ and $m_{\eta_I}^2 = (m_{U_I}^2 + 2m_{S_I}^2)/3$. 

The first important feature of this expression is that there are no free parameters in $B(t)$.  Both the shape and the normalization are completely predicted by \chpt\ as long as one uses point sources for the scalar interpolating field.  The meson masses and the coefficient $\mu$ can be determined from spectrum calculations.  The values of mixed-meson splittings $\Delta_{\rm mix}$ and $\Delta_I$ that appear in $m_{vu}, m_{vs},$ and $m_{U_I},m_{S_I}$ are already known for our choice of valence and sea quark actions.  As a practical matter, we do not even need the computed residual mass as long as we fit $B(t)$ in terms of the meson masses.  

We note that the parameter $\mu$, which relates the quark masses to the squared meson masses at leading order in \chpt, can be determined in multiple ways.  One way is to perform a linear fit to $m_\pi^2$ vs.\ $m_q$.  The resulting value of $\mu\equiv\mu_\textrm{tree}$ will then be common for all valence quark masses at a given lattice spacing.  An alternative method is to define $\mu$ by the following ratio:\be\label{eq:muratio}
	\mu_{\rm ratio} = \frac{m_{\pi,vv}^2}{2(m_v + m_{\rm res})},
\ee
which comes from the tree-level \chpt\ expression for the meson masses.  This gives rise to a different value of $\mu$ for each sea quark ensemble and valence quark mass.  We have calculated $\mu$ using both of these methods, and find that the results are consistent within errors.  We therefore use $\mu_\textrm{ratio}$ for the remainder of our analysis.  

We have considered yet another possibility, which is to take $\mu$ in the chiral limit from a fit to the pseudoscalar mass data including higher order corrections.  By taking $\mu_{\rm ratio}$ (or $\mu_{\rm tree}$), rather than $\mu$ in the chiral limit, one incorporates a subset of higher order corrections to the one-loop formulas.  This leads to better agreement between \chpt\ and numerical data for many purely staggered quantities studied by MILC, including the scalar correlator \cite{Aubin:2004fs, Bernard:2007qf}.  We also find this to be the case in the present work.

The second important feature of the expression for $B(t)$  is that, unlike in continuum full QCD, it receives unphysical contributions from two-pion intermediate states.  If the simulated valence quark masses are sufficiently small, these can dominate the scalar correlator, $C(t)$, at large times and make the $a_0$ mass difficult to determine. 
Because one of the pions in the two-pion bubble is a neutral pion, it has a double pole in its momentum-space propagator in the isospin limit.  The linear-in-$t$ growth factor in the third term of Eq.~(\ref{eq:bubble}) comes from the resulting double pole,
while the minus signs in front of the second and third terms  come from the hairpin residues.

Two-pion intermediate states contribute to $B(t)$ whenever the lattice theory is not unitary; thus they are generically present in any partially-quenched theory.  It is only when both the masses are equal and the actions are identical in the valence and sea sectors that terms 2--4 in Eq.~(\ref{eq:bubble}) vanish and $B(t)$ is strictly positive definite.   Because full QCD is only recovered in the mixed-action theory after taking the continuum limit, two-pion contributions to the scalar correlator cannot be removed completely for \emph{any} choice of mixed-action lattice simulation parameters.  To illustrate this, we consider the two most ``natural" tuning choices for the mixed domain-wall valence, staggered sea theory.  The first possibility is to fix the valence pion mass to be equal to the taste-pseudoscalar sea pion mass.  This tuning is appealing because the mass of the taste-pseudoscalar pion vanishes in the chiral limit, even at finite lattice spacing.  This choice, however, increases the unitarity-violating contributions to $B(t)$ because the taste-singlet pion (the only taste in the sea sector which appears) is much heavier than the taste-pseudoscalar pion on the coarse MILC lattices.  The second possibility is to fix the valence pion mass to be equal to the taste-singlet sea pion mass.  This tuning completely removes the third term in $B(t)$, which has the enhanced (polynomial in $t$) unitarity violation, but still does not eliminate all two-pion contributions.  Furthermore, this tuning may not be advisable in practice since it would lead to a rather heavy valence pion on the coarse MILC lattices.

Thus, although we could in principle tune the mixed-action theory to remove the enhanced unitarity violation in the $a_0$ correlator (and in other quantities such as the $I=0$ $\pi-\pi$ scattering phase shift~\cite{Golterman:2005xa}), the lack of unitarity will always appear to some degree.  The only way to completely remove 
behavior such as negative values 
of the scalar correlator is to take the continuum limit, since both rooted-staggered quarks and domain-wall quarks should reproduce the same continuum theory as $a\to 0$.  For our purposes, however, a large negative bubble contribution to the scalar correlator is actually helpful.  As we will show in the next section, it allows us to easily see the bubble contribution to our data, and therefore to evaluate how well the behavior of the bubble is described by mixed-action chiral perturbation theory.

\subsection{Lattice results for the scalar correlator}\label{sec:scalarlat}

\begin{figure}[htbp]
\begin{center}
\includegraphics[width=4in]{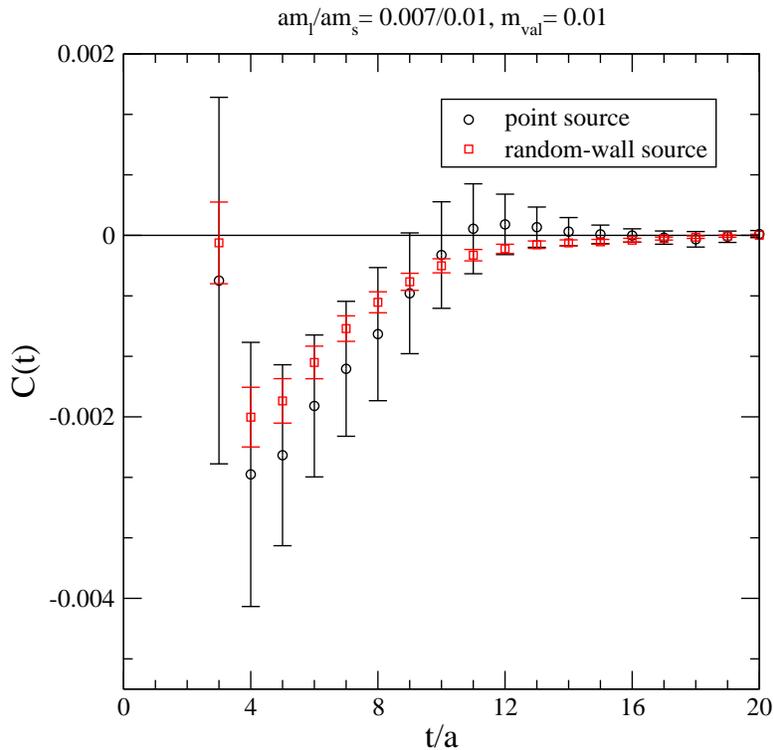}
\caption{Comparison of the scalar correlator generated with a point source on 206 configurations and with a random-wall source on 184 configurations.  The data shown is for $am_v$=0.01 on the 0.007/0.05 coarse ensemble.}
\label{fig:PPvsRW}
\end{center}
\end{figure}

We have calculated the scalar correlator on a subset of the ensembles listed in Table~\ref{tab:MILC_ens}. These are shown in Table~\ref{tab:scalar_ens}, along with the valence quark masses and number of configurations used.  Our preliminary calculation of the scalar correlator presented at Lattice 2007 \cite{Aubin:2007wr} used point sources and sinks because this is the only case for which MA\chpt\ gives a prediction for the bubble term with no unknown parameters.  We found, however, that the statistical errors using the point source were too large to make a compelling case for the agreement between MA\chpt\ and our numerical data, especially on the fine lattices.  This motivated us to switch to a random-wall source, which simulates many point sources and allows us to significantly improve our statistics. The random-wall source has the same normalization as the point source, up to a factor of the spatial volume, so Eq.~(\ref{eq:bubble}) still applies.  Fig.~\ref{fig:PPvsRW} shows a comparison of the point source data with the random-wall source data on the 0.007/0.05 coarse ensemble with a valence domain-wall mass of 0.01 in lattice units.  Although both simulations used approximately 200 configurations (206 for the point and 184 for the random-wall), the statistical errors in the random-wall data are a dramatic factor of 5-6 times smaller than those of the point data.

\begin{table}
\caption{Data  used in the scalar correlator analysis.}
\begin{tabular}{cccc} \\ \hline
$a$(fm) & $a m_l / am_s$ &  $am_v$ & \# configs 
\\[0.5mm] \hline\hline
0.09 & 0.0062/0.0031 & 0.0062 & 80 \\
\hline
0.12 & 0.007/0.05 & 0.01 & 184\\
0.12 & 0.007/0.05 & 0.02 & 175 \\
0.12 & 0.007/0.05 & 0.03 & 189 \\
\hline
\end{tabular}
\label{tab:scalar_ens}
\end{table}

\begin{figure}[htbp]
\begin{center}
\includegraphics[width=4in]{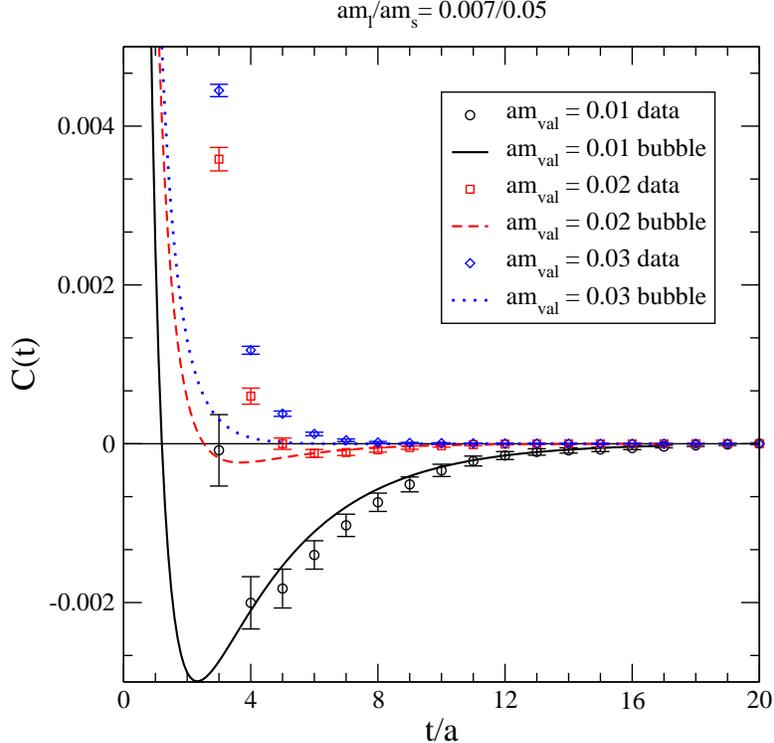}
\caption{Scalar correlator on the coarse 0.007/0.05 ensemble with three different valence masses. Overlaid on the data are the predicted bubble contributions, which should dominate for large time.}
\label{fig:alldata}
\end{center}
\end{figure}

\begin{figure}[htbp]
\begin{center}
\includegraphics[width=4in]{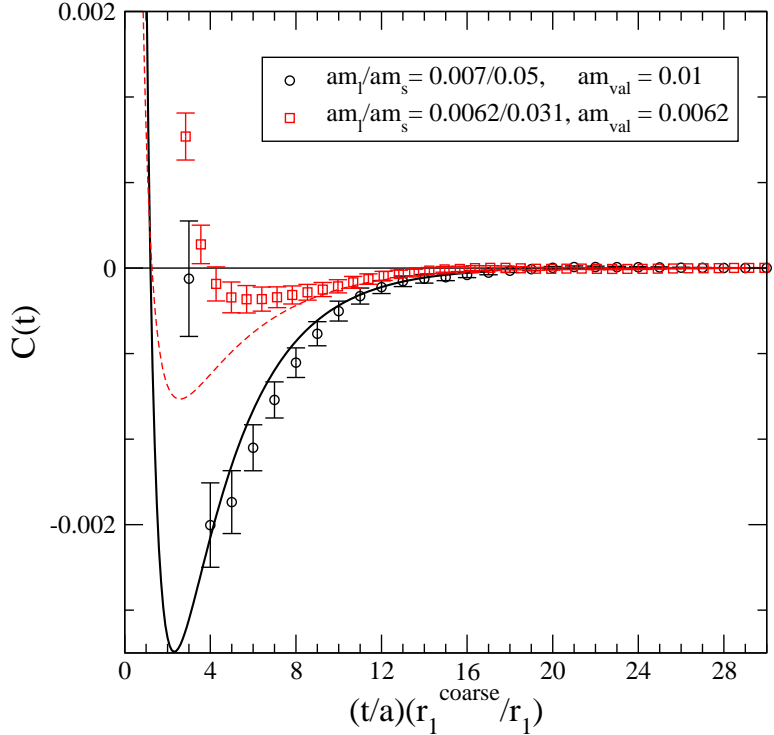}
\caption{Scalar correlator on the coarse 0.007/0.05 ensemble for $am_v$=0.01 and on the fine 0.0062/0.031 ensemble for $am_v$=0.0062. As in Fig.~\ref{fig:alldata}, we include the prediction for the bubble expression for comparison. Note that we have rescaled the time axis for the results on the fine lattice so that we can compare the two different lattice spacings.  The y-axis is dimensionless.}
\label{fig:data_both_as}
\end{center}
\end{figure}

In order to compare our numerical data to MA\chpt,  we first overlay the prediction for the bubble term [Eq.~(\ref{eq:bubble})] on top of the scalar correlator data as a function of time for a number of different masses and two lattice spacings.  Fig.~\ref{fig:alldata} shows three different valence masses on the 0.007/0.05 coarse ensemble, while Fig.~\ref{fig:data_both_as} shows two similar valence masses but on different lattice spacings.  The prediction for the bubble term in Eq.~(\ref{eq:bubble}) requires as inputs three parameters of the chiral effective theory: $\Delta_{\rm mix}$, which we have calculated in Section IV; $\Delta_I$, which we take from Ref.~\cite{Aubin:2004fs}; and $\mu$, which we obtain from our mixed-action pseudoscalar data using Eq.~(\ref{eq:muratio}).  The bubble term is expected to dominate the scalar correlator at large times, and in this region the agreement between the data and the \chpt\ prediction is good.  For sufficiently small times, the direct term $Ae^{-m_{a_0}t}$ dominates, and we would not expect the bubble to match the data in this region.  Note that the trends in the data are qualitatively what one would expect based on Eq.~(\ref{eq:bubble}); Fig.~\ref{fig:alldata} shows that, as the valence quark mass is made lighter, the negative contribution to the bubble increases in magnitude at fixed sea quark mass, and the correlator $C(t)$ becomes negative for much of the time range for the smallest mass.  Fig.~\ref{fig:data_both_as} illustrates the lattice spacing dependence of the scalar correlator and the predicted bubble term. The negative contribution to the bubble is much smaller on the fine lattice than on the coarse even though the valence-valence and taste pseudoscalar sea-sea pion masses are roughly comparable in physical units; this is because the mass-splittings $a^2\Delta_\textrm{mix}$ and $a^2 \Delta_I$ are smaller.

\begin{figure}[htbp]
\begin{center}
\includegraphics[width=4in]{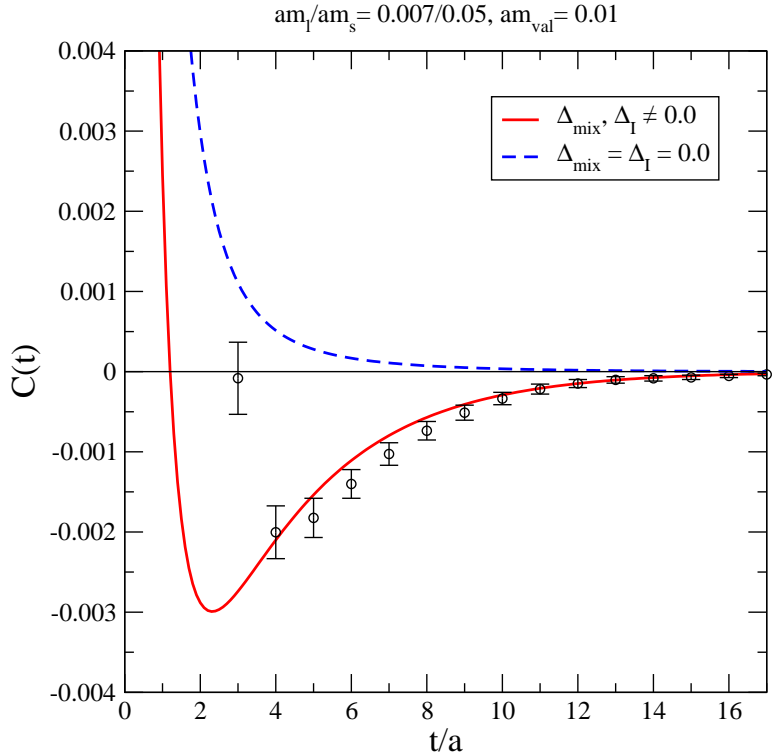}
\caption{Comparison of the bubble prediction for the $am_l/am_s = 0.007/0.05,am_v = 0.01$ mixed-action data with and without the mass-splittings .}
\label{fig:007012bubs}
\end{center}
\end{figure}

If we ignore lattice artifacts and use the continuum form of the bubble expression (by setting $a^2\Delta_I = a^2\Delta_{\rm mix} = 0$), we find that the bubble contribution never becomes negative for the masses used in this work, which is in clear disagreement with the data.  The continuum partially-quenched bubble is plotted in Fig.~\ref{fig:007012bubs} for the parameters corresponding to our lightest valence mass on the coarse lattice, along with the actual numerical data and the bubble prediction including the unitarity-violating lattice artifacts.  The continuum curve does not agree with the data, while the curve that includes the mass-splittings is in good agreement for large times.  This further emphasizes the need to include lattice artifacts such as the mass-splittings in the chiral effective theory in order to describe correctly the lattice data.

\begin{figure}[htbp]
\begin{center}
\includegraphics[width=4in]{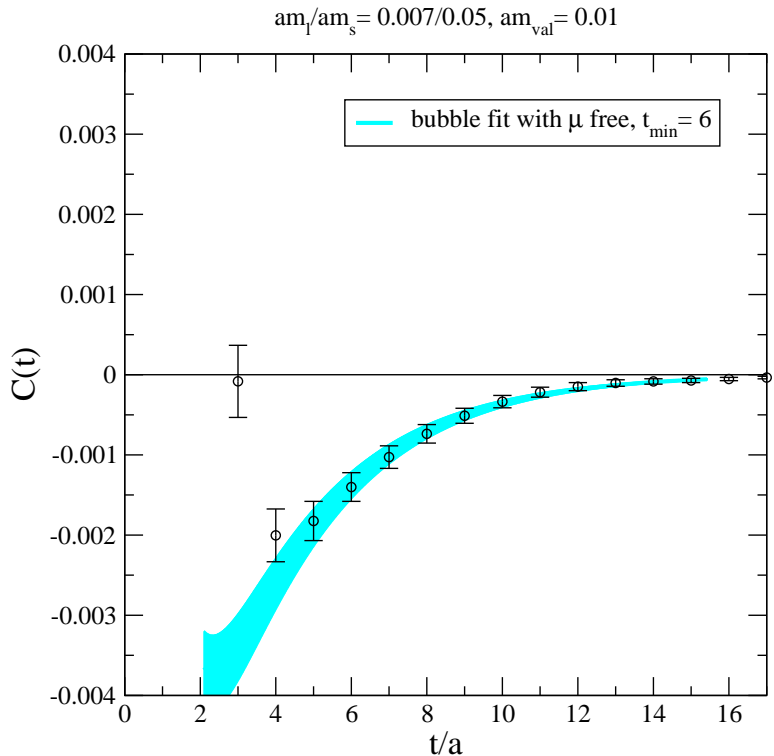}
\caption{Fit of the $am_l/am_s = 0.007/0.05,am_v = 0.01$ data at large times to the bubble expression, $B(t)$, keeping $\mu$ as a free parameter.  The fit range was 6 to 20.  }
\label{fig:00701fits}
\end{center}
\end{figure}

As a consistency check of the chiral parameter $\mu$ that is used in the bubble prediction, we fit the various data sets with $\mu$ free at large times where we can ignore the direct exponential contribution to the scalar correlator.  The results of these fits are shown in Table~\ref{tab:scalarfits2}.  We find that the values for $\mu$ determined from the fits are consistent with the values of $\mu_{\rm ratio}$ used in the prediction of the bubble term at the $ \sim1.5 \sigma$ level.  For the valence mass 0.03 on the coarse ensemble, the bubble term is negligible compared to the statistical errors on the data, and this test is not possible.  Fig.~\ref{fig:00701fits} shows a plot of the scalar correlator for the lightest valence mass on the coarse ensemble with a fit to the bubble expression at large times.
\begin{table}
\begin{tabular}{cccccc}\hline
$am_l$ & $am_v$ & $a\mu_{\rm free}$& $a\mu_{\rm ratio}$ & ($t_{\rm min},t_{\rm max}$)  & $\chi^2/{\rm d.o.f.}$\\
\hline\hline
0.0062 & 0.0062 & 0.86(14) & 1.070 & (12,25) &1.24 \\\hline
0.007 & 0.01    & 1.81(11) & 1.629 & (6,20) &0.98  \\
0.007 & 0.02    & 1.42(33) & 1.541 & (6,20) &1.46  \\
%0.007 & 0.03    & 5.86(34)& 1.491 & (6,20) &1.7 \\
\hline
\end{tabular}
\caption{Fits to $B(t)$ at large times leaving $\mu$ as a free parameter.  For comparison we also show the value of $\mu_{\rm ratio}$ as determined by Eq.~(\ref{eq:muratio}).}
\label{tab:scalarfits2}
\end{table}

\begin{figure}[htbp]
\begin{center}
\includegraphics[width=4in]{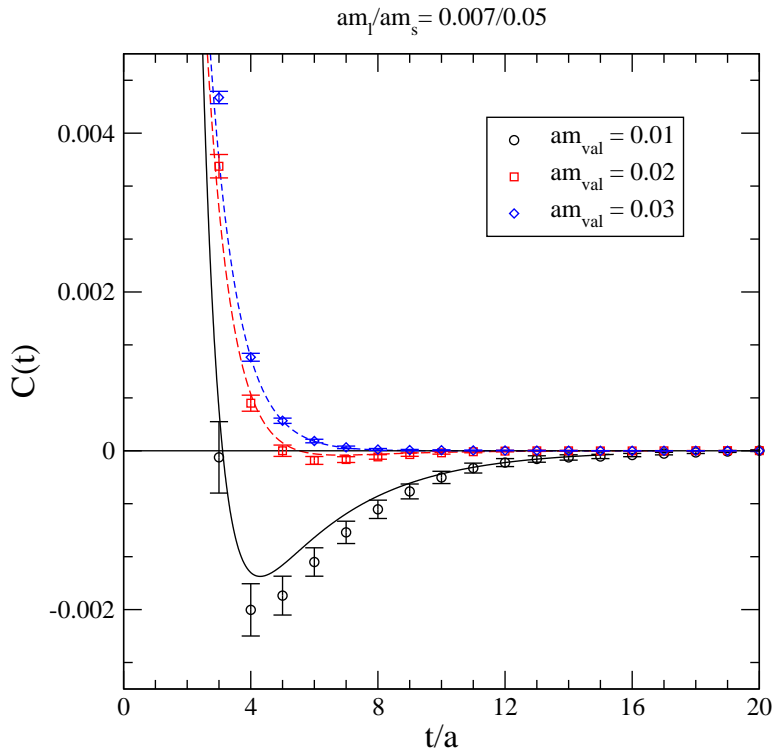}
\caption{Fits to all three valence quark masses on the $0.007/0.05$ coarse ensemble.  The fit function is given in Eq.~(\ref{eq:corr}), and contains a single exponential plus the bubble contribution with $\mu=\mu_{\rm ratio}$.}
\label{fig:allcoarsefits}
\end{center}
\end{figure}

\begin{figure}[htbp]
\begin{center}
\includegraphics[width=4in]{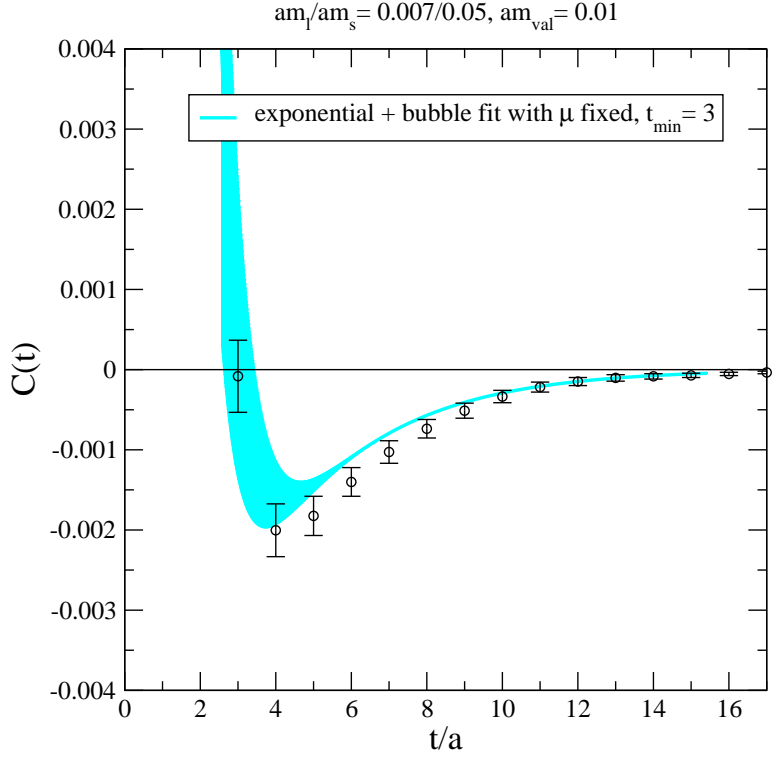}
\caption{Fit to the $am_l/am_s$=0.007/0.05, $am_v$=0.01 data with error band displayed.  Note that this is the same fit as the solid black curve in Fig.~\ref{fig:allcoarsefits}.}
\label{fig:00701fitratio}
\end{center}
\end{figure}

\begin{figure}[htbp]
\begin{center}
\includegraphics[width=4in]{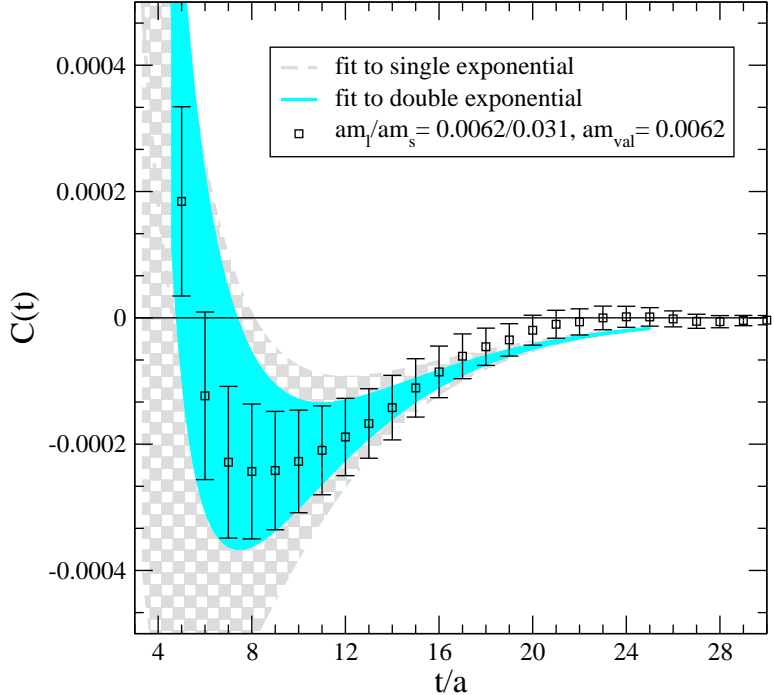}
\caption{Comparison of fits to the fine scalar data with a single (grey checked) or double (cyan solid) exponential term.  The single exponential fit required a $t_{min}$ of 7 in order to obtain a reasonable $\chi^2/\textrm{d.o.f.}$; the double exponential fit allowed a $t_{min}$ of 2.}
\label{fig:00620062fits}
\end{center}
\end{figure}

Having confirmed that the results for $\mu$ from fits to the scalar correlator are consistent with the values of $\mu_{\rm ratio}$ determined from pseudoscalar data, we next perform fits to the full expression for $C(t)$ in Eq.~(\ref{eq:corr}).  Table~\ref{tab:scalarfits1} shows the results of the different fits for each of the ensembles, including the bubble term with $\mu$ set to $\mu_{\rm ratio}$ plus a ground state exponential term.  In order to describe the fine lattice data at small times, we find that we must include an additional exponential term to model excited state contributions. We can get reasonable fits to all masses shown, although we point out that for the heavier mass of $m_v=0.03$ on the coarse lattice, the bubble contribution has a very small effect [for $t\ge 4$, $B(t)\sim 0$], and similar $\chi^2$ results can be obtained while omitting it. For the lighter masses, including the bubble is absolutely necessary.  Figs.~\ref{fig:allcoarsefits} and \ref{fig:00701fitratio} show fits of the coarse data to the full expression for $C(t)$ including only the ground-state exponential plus the bubble term. Fig.~\ref{fig:allcoarsefits} shows the results for all three valence masses on the coarse ensemble, while Fig.~\ref{fig:00701fitratio} shows only the lightest valence mass, but Fig.~\ref{fig:00701fitratio} includes the error band for the resulting fit.  Given our current statistics, we can successfully fit all of the coarse data starting at a minimum time of three or four while including only a single exponential term.   This is not the case, however, for our fine lattice data which has both a smaller bubble contribution and a finer resolution in the time direction.  When we include an excited state term in the fit to the fine lattice data, we are able to obtain an acceptable $\chi^2/\textrm{d.o.f.}$ over a much larger fit range, and we reduce the statistical errors in the fitted parameters, as illustrated by Fig.~\ref{fig:00620062fits}. 

The different fits of the fine lattice data yield consistent values for the ground state meson mass, and we find $am_{a_0}\approx 0.38(5)(3)$, where the first error is statistical and the second is a systematic error which reflects the differing central values between the single and double exponential fits.  Given that $a^{-1}\sim 2.3$ GeV on the fine lattice, we note that this result is close to the experimentally measured $a_0$ meson mass of $\sim$ 980 MeV.  We emphasize, however, that this result is for unphysical quark masses at a nonzero lattice spacing, and that the systematic error does not include other important sources of uncertainty such as quark mass dependence and discretization effects. 

\begin{table}
\begin{tabular}{ccccc}\hline
\# Exp & $am_l$ & $am_v$ & ($t_{\rm min},t_{\rm max}$)  &$\chi^2/{\rm d.o.f.}$\\
\hline\hline
1 & 0.0062 & 0.0062 & (7,24) &1.10  \\
2 & 0.0062 & 0.0062 & (2,21) & 0.87\\\hline
1 & 0.007 & 0.01 & (3,20)& 1.04  \\
1 & 0.007 & 0.02 &(4,20)&0.89  \\
1 & 0.007 & 0.03 &(4,20)&0.57  \\\hline
\end{tabular}
\caption{Fits to the mixed-action scalar data using the expression for $C(t)$ given in Eq.~(\ref{eq:corr}), keeping $\mu$ fixed to the ratio value defined in Eq.(\ref{eq:muratio}).  For the fine lattice data only, we show results for both one and two exponential terms in the fit function.  We do not show the resulting masses and amplitudes in this table because our purpose is not to calculate the $a_0$ mass, but rather to demonstrate that we can successfully fit the data with a physically motivated fitting function.}
\label{tab:scalarfits1}
\end{table}

Although it has been suggested in Ref. \cite{Prelovsek:2005rf} that the scalar correlator could be used to determine $\Delta_{\rm mix}$, we have found this to be very difficult with the parameters and level of statistics used in the present work.  We note that varying $\Delta_{\rm mix}$ within the statistical errors from Sec.~\ref{sec:DelMix}, or even by a factor of two, does not have a significant effect on the numerical value of the bubble term for our values of the simulation parameters. Consequently, since the bubble term is not very sensitive to this variation, we would not expect $\Delta_{\rm mix}$ to be well-constrained by our scalar data, and fits leaving $\Delta_{\rm mix}$ as a free parameter confirm this expectation.  Finally, we observe that the bubble term is much more sensitive to $\Delta_I$, and the consistency of our results with MA\chpt\ provides a good check of the MILC determination of $\Delta_I$.

We conclude this section by observing that our data is quite well described by the mixed-action \chpt\ formula calculated by Prelovsek. This highly non-trivial test indicates that mixed-action \chpt\ can describe the largest source of unitarity violations that appear in our simulations. Additionally, our data shows that the degree of unitarity violation decreases with the lattice spacing, as expected.  We note, however, that other sources of unitarity violations are expected to be present at some level, although we cannot currently resolve them in our data.  These are due to short-distance effects and therefore cannot be described by \chpt.  The first of these effects is due to the fact that domain-wall fermions do not have a positive-definite transfer matrix \cite{Furman:1994ky}.  This leads to the appearance of opposite parity states in two-point correlators, which are similar to typical excited states, except that they oscillate in sign from one time slice to the next.  Evidence of these oscillating states using the same mixed-action scheme as this work was presented in Ref.~\cite{Syritsyn:2007mp}, where the effect was shown to be small in the pion two-point correlator.  We note that oscillating states also contribute in the staggered valence on staggered sea case, and the analysis of the scalar correlator in Ref.~\cite{Bernard:2007qf} required such terms in order to describe the numerical data.  We do not see evidence of oscillating states, however, in our own scalar correlator data.  The second of the unitarity-violating effects is the presence of enhanced zero-mode contributions in the (partially) quenched theory; this was studied in Ref.~\cite{Blum:2000kn}.  These effects are known to be suppressed by the fermion determinant, large spatial volumes, and large time separations.  Although such effects must be present in partially-quenched domain-wall correlators at some level, we see no evidence for them in this work.  As we have shown, our numerical mixed-action scalar data is quantitatively consistent with the MA\chpt\ bubble prediction plus a small number of non-oscillating exponential terms.

%=================================================
\section{Conclusions}
\label{sec:Conc}
%=================================================

Mixed-action lattice simulations using domain-wall valence quarks and improved staggered sea quarks are well-suited to the calculation of weak matrix elements such as $B_K$.  Discretization effects in the mixed-action theory arise from both the valence and sea sectors, as well as the interaction between the two.  Although such effects vanish in the continuum limit, in practice they must be removed from quantities calculated at finite lattice spacing with the aid of extrapolation formulae calculated in mixed-action Chiral Perturbation Theory.  

In this work we have studied the size of discretization effects in the LHP Collaboration mixed-action scheme which uses the MILC Asqtad-improved staggered lattice configurations.  We have calculated the values of the residual quark mass, $m_\textrm{res}$, and the mixed valence-sea meson mass-splitting, $\Delta_\textrm{mix}$, and have found them to be comparable in size to or smaller than analogous parameters in numerical simulations by the RBC, UKQCD, and MILC Collaborations.  Thus we conclude that the size of generic discretization effects in the mixed-action theory are small enough to allow precise determinations of continuum weak matrix elements.  We have also performed a strong check of the ability of MA\chpt\ to accurately describe unitarity-violating discretization effects in the isovector scalar correlator.  We find that the MA\chpt\ prediction for the two-particle intermediate state (bubble) contribution to the scalar correlator is in good quantitative agreement with the numerical lattice data.  Thus we conclude that MA\chpt\ correctly describes the dominant unitarity-violating contributions to mixed-action lattice simulations.  

In the case of the scalar meson, the unitarity-violating discretization effects are much larger than the  $a_0$ meson ground-state signal for many of our quark mass values, so any uncertainty in the bubble contribution translates directly into an uncertainty in the extracted scalar mass.  One may therefore argue that mixed-action (or any partially-quenched) simulations cannot be used for a clean determination of the scalar meson mass.   It was shown in Ref.~\cite{Prelovsek:2004jp}, however, that for the case of  $N_f=2$ domain-wall fermions, use of partially-quenched data allows a more precise extraction of the $a_0$ mass than with full QCD points alone, despite contamination from unphysical intermediate states.    In fact, it is interesting to note that, because MA\chpt\ predicts the observed unitarity violations so well, one can use the generic framework of partially-quenched lattice \chpt\ to aid in choosing the best actions and parameters for determining the $a_0$ mass.  We have not done this because our primary goal is the calculation of weak matrix elements.  Fortunately, in the case of $B_K$, MA\chpt\ predicts that non-analytic unitarity-violating errors should contribute less than a percent on the coarse MILC lattices.  This fact, in conjunction with our successful analysis of the scalar correlator, substantiates the claim that unitarity-violating effects in mixed-action lattice simulations can be accounted for and removed to recover precise continuum values for weak matrix elements.   

%=================================================
\section*{Acknowledgments}
%=================================================

Computations for this work were carried out on facilities of the USQCD Collaboration, which are funded by the Office of Science of the U.S. Department of Energy.  We thank Huey-Wen Lin and Meifeng Lin for help with the Columbia Physics System, Robert Edwards and Balint Jo\'o for help with Chroma, and Jim Simone for help with FermiQCD and the MILC code.   We thank Oliver B\"ar and Sasa Prelovsek for valuable discussions, and Claude Bernard and Steve Sharpe for useful comments on the manuscript.  Christopher Aubin is supported by the DOE under grant nos. DE-FG02-92ER40699 and DE-FG02-04ER41302.   Jack Laiho is supported by the DOE under grant DE-FG02-91ER40628 and by the NSF under grant PHY-0555235.  Fermilab is operated by Fermi Research Alliance, LLC  under Contract No. DE-AC02-07CH11359 with the United States Department of Energy.  

%=================================================
%The bibliography
%=================================================

\bibliography{SuperBib}

\end{document}